\documentclass[pdflatex,sn-chicago]{sn-jnl}

\usepackage{graphicx}%
\usepackage{multirow}%
\usepackage{amsmath,amssymb,amsfonts}%
\usepackage{amsthm}%
\usepackage{mathrsfs}%
\usepackage[title]{appendix}%
\usepackage{xcolor}%
\usepackage{textcomp}%
\usepackage{manyfoot}%
\usepackage{booktabs}%
\usepackage{algorithm}%
\usepackage{algorithmicx}%
\usepackage{algpseudocode}%
\usepackage{listings}%
\usepackage{tikz}
\usepackage{pgfplots}
\pgfplotsset{compat=1.12}
\usepgfplotslibrary{fillbetween}
\usetikzlibrary{patterns,backgrounds,fit,shapes.geometric}
\usetikzlibrary{arrows.meta,
                positioning,
                shadows}
\usepackage{soul}

\tikzset{
    many copy shadow/.style={
        copy shadow={shadow xshift=12pt, shadow yshift=12pt},
        copy shadow={shadow xshift=10pt, shadow yshift=10pt},
        copy shadow={shadow xshift=8pt, shadow yshift=8pt},
        copy shadow={shadow xshift=6pt, shadow yshift=6pt},
        copy shadow={shadow xshift=4pt, shadow yshift=4pt},
        copy shadow={shadow xshift=2pt, shadow yshift=2pt} 
    }
}

\tikzset{
    some copy shadow/.style={
        copy shadow={shadow xshift=8pt, shadow yshift=8pt},
        copy shadow={shadow xshift=6pt, shadow yshift=6pt},
        copy shadow={shadow xshift=4pt, shadow yshift=4pt},
        copy shadow={shadow xshift=2pt, shadow yshift=2pt} 
    }
}

\tikzset{
    two copy shadow/.style={
        copy shadow={shadow xshift=4pt, shadow yshift=4pt},
        copy shadow={shadow xshift=2pt, shadow yshift=2pt} 
    }
}

\usepackage{natbib}
\setcitestyle{authoryear,open={(},close={)}}

\theoremstyle{thmstyleone}%
\theoremstyle{thmstyletwo}%

\theoremstyle{thmstylethree}%

\begin{document}

\title[Article Title]{Bidding in Ancillary Service Markets: An Analytical Approach Using Extreme Value Theory}


\author[1]{\fnm{Torine} \sur{Reed Herstad}}\email{torhe@dtu.dk}

\author[1]{\fnm{Jalal} \sur{Kazempour}}\email{jalal@dtu.dk}

\author[1]{\fnm{Lesia} \sur{Mitridati}}\email{lemitri@dtu.dk}

\author[2]{\fnm{Bert} \sur{Zwart}}\email{bert.zwart@cwi.nl}

\affil[1]{\orgdiv{Department Wind and Energy Systems}, \orgname{Technical University of Denmark}, \orgaddress{\street{Elektrovej}, \city{Kgs. Lyngby}, \country{Denmark}}}

\affil[2]{\orgname{CWI, National Research Institute of Mathematics and Computer Science}, \orgaddress{\city{Amsterdam}, \country{The Netherlands}}}


\abstract{To enable the participation of stochastic distributed energy resources in ancillary service markets, the Danish transmission system operator, Energinet, mandates that flexibility providers satisfy a minimum 90\% reliability requirement for reserve bids. This paper examines the bidding strategy of an electric vehicle aggregator under this regulation and develops a chance-constrained optimization model. In contrast to conventional sample-based approaches that demand large datasets to capture uncertainty, we propose an analytical reformulation that leverages extreme value theory to characterize the tail behavior of flexibility distributions. A case study with real-world charging data from 1400 residential electric vehicles in Denmark demonstrates that the analytical solution improves out-of-sample reliability, reducing bid violation rates by up to 8\% relative to a sample-based benchmark. The method is also computationally more efficient, solving optimization problems up to 4.8 times faster while requiring substantially fewer samples to ensure compliance. Moreover, the proposed approach enables the construction of feasible bids with reliability levels as high as 99.95\%, which would otherwise require prohibitively large scenario sets under the sample-based method. Beyond its computational and reliability advantages, the framework also provides actionable insights into how reliability thresholds influence aggregator bidding behavior and market participation. This study establishes a regulation-compliant, tractable, and risk-aware bidding methodology for stochastic flexibility aggregators, enhancing both market efficiency and power system security.}





\maketitle
\bmhead{Acknowledgment}
The authors would like to thank Spirii for providing the data, as well as Gustav A. Lunde and Emil V. Damm for their extensive work on data preparation, which made this case study possible. The authors would also like to thank the editor and the two anonymous reviewers for their constructive feedback, which has greatly improved the quality of this paper.

\section{Introduction}

\subsection{Background and motivation}

Flexibility aggregators pool and manage distributed energy resources such as electric vehicles (EVs), enabling them to provide ancillary services like frequency regulation and load balancing \citep{Energinet1}. By doing so, they contribute to power system stability while creating new revenue streams for consumers. However, the inherently stochastic nature of distributed resources, arising from uncertain consumption, charging, and availability patterns, makes it difficult to guarantee reliable service delivery. For system operators, ensuring that contracted reserves are available during activation events is essential to maintaining grid security.

To address this challenge, the Danish transmission system operator, Energinet, has introduced grid codes that require at least 90\% reliability for ancillary service bids \citep{Energinet}. This requirement, known as the \textit{P90 requirement}, is particularly relevant for EV aggregators, who must balance uncertain user behavior with regulatory compliance in order to participate in the Nordic market for ancillary services. While the requirement facilitates the market participation of stochastic flexibility providers, it also raises important methodological questions. The key challenge for aggregators is to design regulation-compliant bidding strategies that explicitly manage uncertainty without being either overly conservative or computationally intractable.

Existing optimization techniques provide partial answers but remain limited. Robust optimization guarantees feasibility under worst-case realizations, but the resulting bids are often too conservative, leading to under utilization of available flexibility. Distributionally robust optimization relaxes this conservatism by considering ambiguity sets, yet its effectiveness depends on assumptions about distributional properties that may not hold in practice. Scenario-based and sample-based approaches are widely applied due to their simplicity, but they require very large numbers of samples to approximate low-probability high-impact events, which leads to significant computational costs. More critically, these methods fail to capture extreme shortfalls that drive Energinet’s reliability checks, underscoring the need for tractable bidding models that explicit target tail risks in flexibility estimation.

\subsection{Literature review}

A growing body of research has examined bidding strategies for aggregators of flexible resources under uncertainty. Robust optimization approaches, such as those developed by \citet{kuhn}, directly encode reliability requirements but often yield highly conservative solutions. \citet{GadeP90} propose a bidding strategy for EV aggregators using distributionally robust optimization, aiming to balance tractability and robustness, though it relies heavily on assumptions regarding the ambiguity sets. More recent studies incorporate chance constraints and conditional value-at-risk into aggregator models \citep{GandE}, yet these methods typically rely on sample approximations and therefore inherit the scalability and accuracy limitations of scenario-based approaches.

Scenario-based optimization remains the most commonly adopted method, representing uncertainty through a finite set of samples. While conceptually straightforward, it often requires hundreds or thousands of samples to capture extreme realizations, substantially increasing computational costs \citep*{BertScaling, blanchet2024efficient}. 
Approximation techniques, such as Bonferroni-type bounds or convex reformulations \citep{Prekopa1970, NemirovskiStewart2012}, improve tractability but generally fail to capture the “tail” behavior of uncertainty distributions, where rare yet extreme shortfalls occur.

A fundamental limitation emerges across these studies: each method resolves part of the challenge, but none reconciles the trade-off between over-conservatism, restrictive assumptions, scalability, and accurately capturing tail risks. Extreme value theory offers a pathway to accurately model extreme events while preserving tractability \citep{diebold2000pitfalls}. Traditional tail and extreme value analysis focuses on post hoc risk assessment \citep{husler2003optimization}. Yet, in energy systems, where extreme deviations in supply or demand can have catastrophic impacts on system stability, there is a strong motivation to embed such analysis within the optimization process itself. This presents an opportunity to integrate concepts of \textit{extreme value theory} into chance-constrained optimization, enabling decision-making under extreme uncertainties without the need for large scenario sets. In this direction, \citet{RareCCsTong} combine large deviation theory with convex analysis and bilevel optimization to derive tractable formulations solvable with standard solvers, assuming Gaussian mixture models (GMMs). The inherently light-tailed structure of GMMs presents an important limitation for concrete power system applications, potentially underestimating infrequent but severe flexibility deviations. In contrast, extreme value theory is explicitly designed to model such rare events, providing a more reliable foundation for chance-constrained formulations in this context.

To the best of our knowledge, this paper is the first to embed concepts from extreme value theory within chance-constrained optimization and demonstrate their value in practice by providing a bidding method for aggregators of stochastic flexible resources in ancillary service markets, grounded in real data to capture tail behavior, while also deriving actionable insights on the role and impact of market regulations.

\subsection{Contributions}


This paper addresses these gaps by developing an analytical approach to aggregator bidding that integrates extreme value theory into a chance constrained optimization framework. The main contributions are summarized as follows:

\begin{itemize}
    \item A chance constrained bidding model for an electric vehicle aggregator in Nordic ancillary service markets that directly incorporates Energinet’s 90\% reliability requirement together with the limited energy reservoir constraint.
    \item An analytical reformulation based on extreme value theory that focuses explicitly on the tail of flexibility distributions, enabling tractable solutions with far fewer samples than conventional scenario-based methods.
    \item Numerical validation with real-world data from 1400 residential charging points in Denmark, showing that the extreme value theory based approach reduces out-of-sample violation rates by up to 8\% and solves optimization problems up to 4.8 times faster than a sample-based benchmark.
    \item A demonstration of stricter reliability compliance, where the proposed model produces feasible bids at reliability levels as high as 99.95\%, which would require prohibitively large scenario sets under a sample-based approach.
    \item Practical insights for aggregators and system operators, showing how changes in reliability thresholds influence bidding strategies, revenue opportunities, and market participation.
\end{itemize}

Together, these contributions establish a novel, regulation-compliant, and computationally efficient framework for stochastic flexibility bidding, enhancing both market efficiency and power system security.

\subsection{Paper organization}

The remainder of this paper is structured as follows. Section \ref{sec:preliminaries} introduces the necessary preliminaries for understanding the subsequent sections. Section \ref{sec:m&m} presents the formulation of the chance-constrained optimization problem, while Section \ref{sec:analyticaltheory} outlines the methodology for its analytical reformulation, detailed further in Section \ref{sec:analyticalsolution}. Section \ref{sec:SAAform} describes a commonly used sample-based approach for comparison. Numerical results are reported in Section \ref{sec:results}, and Section \ref{sec:concandfuture} concludes the paper. Additional details on the case study and the distribution fitting procedure are provided in Appendices A and B.


\section{Preliminaries}\label{sec:preliminaries}
This section provides an overview of Nordic ancillary service markets and the relevant grid codes for stochastic distributed energy resources participating in these markets.

\subsection{FCR-D market}
Frequency containment reserve (FCR) is a category of ancillary services that helps balance short-term deviations in power and maintain the frequency stability of the electrical grid. In the Nordic grid, known for its low inertia, relatively small system size, and high share of stochastic renewable energy, FCR is divided into multiple services. These include FCR for normal operation (FCR-N), which is activated when the frequency deviates between 49.9 Hz and 50.1 Hz, and FCR for disturbances (FCR-D), which is activated when frequency deviations exceed the normal operational range. Specifically, FCR-D up service is triggered when the frequency drops below 49.9 Hz, while FCR-D down service is activated when the frequency rises above 50.1 Hz. To stabilize the grid, FCR-N and FCR-D service providers adjust their power production or consumption levels in response to frequency fluctuations. FCR-N is a symmetrical product, with equal bids made in both directions, whereas FCR-D is asymmetrical and split into two separate products: upward and downward \citep{Energinet}. FCR-N and FCR-D are procured in separate day-ahead markets, each tailored to their specific operational roles. Without loss of generality, this paper focuses on the FCR-D market, since its asymmetric bidding structure for upward and downward services aligns well with the characteristics of EV aggregators and their battery state-of-charge constraints.

\subsection{The P90 requirement}
To maintain grid stability, the provision of ancillary services must be highly reliable. However, imposing a 100\% reliability requirement for ancillary service bids would limit the participation of many non-conventional technologies, such as wind power or aggregated flexible loads like EVs. To facilitate the participation of stochastic flexible resources in  ancillary service markets, including FCR-D, which is the focus of this paper, Energinet has introduced the P90 requirement, as outlined in \citeauthor{Energinet} \citeyearpar{Energinet} and \citeauthor{GandE} \citeyearpar{GandE}.

The P90 requirement ensures that flexibility providers, such as EV aggregators, can participate in  ancillary service markets, provided that their reserved capacity bid is fully realized at least 90\% of the time. To be eligible to bid into the markets, an aggregator must complete Energinet's prequalification process \citep{Energinet} and conduct a prognosis analysis of its ability to meet future demand, which is verified by Energinet. Once qualified by Energinet, a flexibility provider can participate in the relevant ancillary service markets. From that point, Energinet continuously performs \textit{ex-post} checks to verify whether a reserved capacity bid by the participant is truly available, regardless of whether the reserve is activated. This serves as an additional security measure for supply.  
For example, an FCR-D up-reserve bid of 10 kW becomes unavailable if the actual consumption of an EV aggregator is only 8 kW, since the aggregator cannot reduce its consumption by the full 10 kW. Any partial or full unavailability of a bid counts as a \textit{reserve shortfall}. Energinet conducts these checks at appropriate time intervals, such as every yearly quarter. If a provider violates the P90 requirement, meaning that the frequency of non-zero shortfalls exceeds 10\% of the time, it will be excluded from further market participation and must re-apply for qualification.

In the context of modelling such requirements mathematically, this requirement can be expressed using chance constraints of the form $\mathbb{P}(\cdot) \geq 1 - \varepsilon$, where $\varepsilon=0.1$ represents the maximum 10\% allowance for failure in successfully delivering the service.

\subsection{The LER requirement}
In addition to the P90 requirement, Energinet has introduced an additional requirement for resources with limited energy reservoirs (LER), such as batteries and EV aggregators. The primary objective of the LER requirement is to ensure that resources with limited energy capacity can maintain energy availability despite their constraints. This includes adhering to energy management protocols and ensuring the ability to sustain continuous activation for a specified minimum duration when needed. The definition of this requirement is outlined in \citeauthor{Energinet} \citeyearpar{Energinet} and \citeauthor{GandE} \citeyearpar{GandE}.

The specifics of the LER requirement differ between FCR-N and FCR-D services. This paper focuses on FCR-D, where an LER unit bidding in one direction of the market (either up or down) must also ensure that at least 20\% of its capacity is available in the opposite direction. Mathematically, this is captured in \eqref{eq:1a}, which requires reserving 20\% of the available upward flexibility for downward activation. For EV aggregators, this LER requirement applies only in the upward direction, because downward activation (i.e., increasing power consumption) is limited by the battery storage capacity of the EV. In contrast, upward activation (i.e., reducing power consumption) is not constrained by battery capacity since the power supply to a charging EV can always be turned off and is treated like other non-LER flexible loads. The requirement further specifies that the service must remain available for at least 20 continuous minutes once activated.


\begin{figure}
  \centering
  \begin{tabular}{@{}c@{}}
    \includegraphics[width=.7\linewidth]{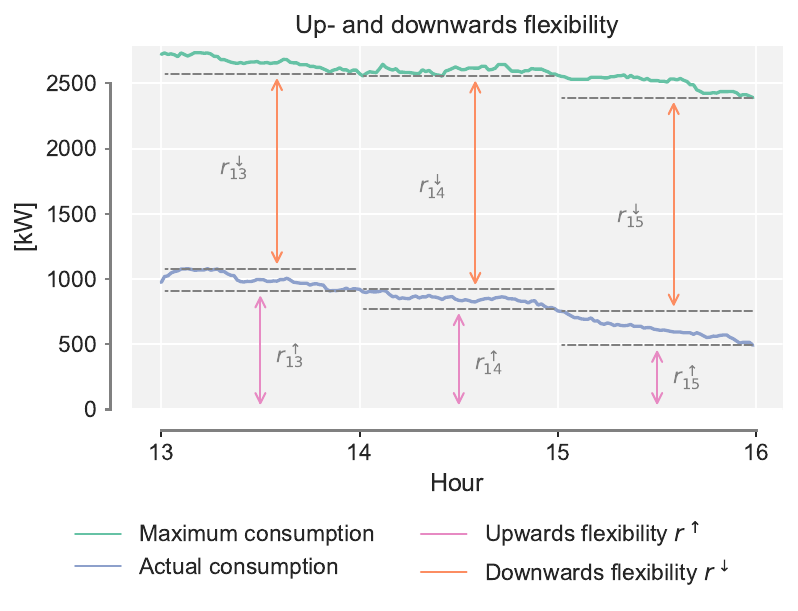} \\[\abovecaptionskip]
    \footnotesize (a) Downwards and upwards flexibility for an aggregated EV fleet.
  \end{tabular}

  \vspace{\floatsep}

  \begin{tabular}{@{}c@{}}
    \includegraphics[width=\linewidth]{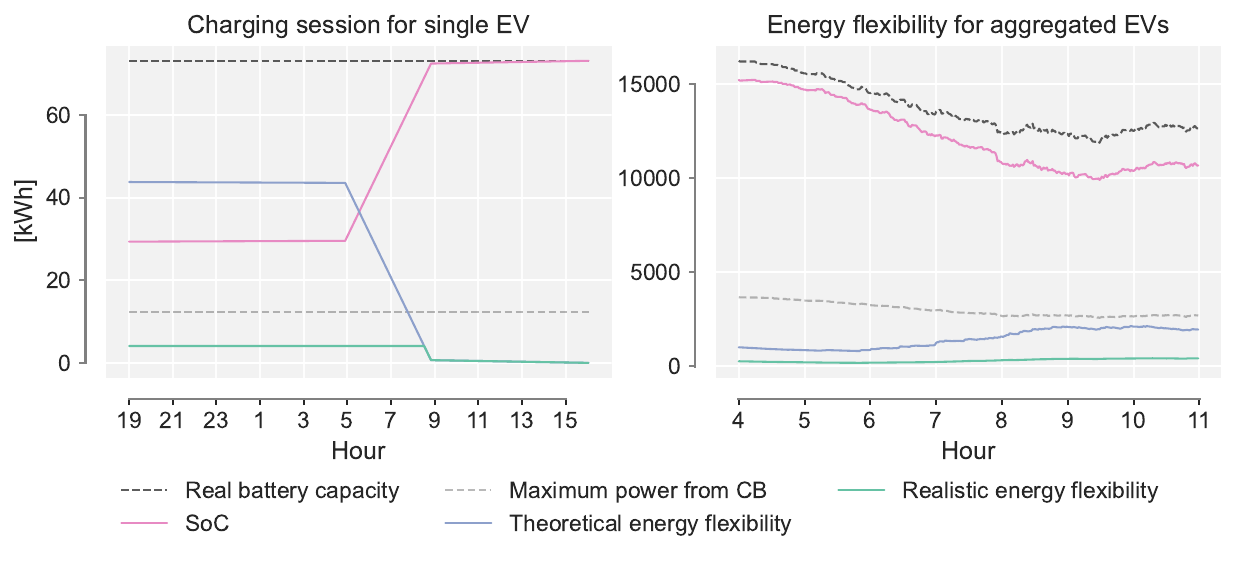} \\[\abovecaptionskip]
    \footnotesize (b) Energy flexibility for a single EV and for an aggregated fleet.
  \end{tabular}

  \caption{Illustrative examples of flexibility estimation: Figure (a) illustrates how downwards and upwards flexibility for the aggregated case are calculated. Downwards flexibility ($r^\downarrow$) is determined as the difference between the lowest point of the maximum consumption of the EVs and the highest actual consumption across the hour, while upwards flexibility ($r^\uparrow$) is the minimum of the actual consumption values during the hour. The figures in (b) show how energy flexibility ($r^{\rm{\rm{E_{20}}}}$) is calculated for a single EV and for the aggregated case. The single EV case represents a complete charging session, while the aggregated case captures a time interval in the day with large fluctuations in available flexibility, typically during the morning when many charging sessions end. Energy flexibility is calculated based on the current state of charge (SoC) and the amount of power that can be applied to the EV over the next 20 minutes without exceeding the battery's capacity. For instance, at hour 1 in the single EV case, the SoC is 30 kWh, and the battery capacity is 75 kWh, meaning that, theoretically, the charge box (CB) could apply a maximum of 45 kWh in the next 20 minutes to fully charge the EV. However, due to the CB's physical constraint of 12 kW, the actual available energy flexibility is much lower. Similarly, in the aggregated case, the theoretical and realistic energy flexibilities are on vastly different scales, reflecting the system's capacity and constraints.}\label{fig:flexcalculations}
\end{figure}


These conditions impose the following constraints on the upwards and downwards reserve capacity bids of the EV aggregator (in kW), denoted as $b^\uparrow$ and $b^\downarrow$, which are submitted to the FCR-D market:  
\begin{subequations}\label{eq:LERconstraints}
    \begin{align}
            0.2b^\downarrow + b^\uparrow &\leq R^\uparrow, \label{eq:1a} \\
            b^\downarrow &\leq R^\downarrow, \label{eq:1b} \\
            b^\downarrow &\leq R^{\rm{E_{20}}},\label{eq:1c}
    \end{align}
\end{subequations}
where ${R^\uparrow, R^\downarrow, R^{\rm{E_{20}}}}$ are random variables representing the available upward, downward, and energy flexibility, respectively. Specifically, upward flexibility $R^\uparrow$ in \eqref{eq:1a} refers to the potential for reducing consumption of the grid-connected EV fleet and is subject to the LER requirement described above. Downward flexibility $R^\downarrow$ in \eqref{eq:1b} refers to the potential for increasing consumption. Energy flexibility $R^{\rm{E_{20}}}$ in \eqref{eq:1c} is determined by the current state of charge of the entire EV fleet and the maximum power that can be applied over the next 20 minutes without exceeding battery capacity.

The realizations of these random variables are denoted as $\{r^\uparrow, r^\downarrow, r^{\rm{E_{20}}}\}$. A detailed explanation of how these realizations are defined can be found in Appendix \ref{sec:case study intro}. For a quicker overview, Figure \ref{fig:flexcalculations} provides graphical representations of how the flexibilities are estimated. Specifically, the realizations $\{r^\uparrow, r^\downarrow, r^{\rm{E_{20}}}\}$ are defined as the extreme values of flexibility within the hour, which in this study corresponds to the minimum available flexibility during the hour. In contrast to \citet{GandE}, who define similar sets using minute-level realizations, our hourly approach reduces computational burden by decreasing the number of data points by a factor of 60 while also mitigating temporal dependencies in the time series.


\section{Chance-constrained bidding model}\label{sec:m&m}

To incorporate the requirements imposed by Energinet into the bidding strategy, we employ chance-constrained optimization. This approach manages uncertainty by ensuring that certain constraints are satisfied with a specified probability rather than absolute certainty. For EV aggregators participating in ancillary service markets, chance constraints enable bids that provide a balance between reliability and resource utilization. Specifically, Energinet's grid code mandates that aggregators commit to a service provision that is fully available at least 90\% of the time, allowing for a 10\% margin of non-compliance due to uncertainties in EV availability. 

In line with the P90 and LER requirements, we adopt a joint chance-constrained optimization model similar to the one proposed in \citeauthor{GandE} \citeyearpar{GandE}:
\begingroup
\allowdisplaybreaks
\begin{subequations}\label{eq:MAIN opt form}
    \begin{align}
        &\max_{b^\downarrow, b^\uparrow \geq 0} b^\uparrow + b^\downarrow \label{eq:CC1} \\
        \text{s.t.}\quad& \nonumber \\
        &\mathbb{P}\left(\begin{aligned}
            &0.2b^\downarrow + b^\uparrow \leq R^\uparrow \\
            &b^\downarrow \leq R^\downarrow \\
            &b^\downarrow \leq R^{\rm{E_{20}}}
        \end{aligned} \right) \geq 1 - \varepsilon, \label{eq:CC2}
    \end{align}
\end{subequations}
\endgroup
The objective function in \eqref{eq:CC1} maximizes the combined reserve capacity bids in the FCR-D up and down markets. Constraint \eqref{eq:CC2} defines the joint chance constraint, which incorporates both the LER and P90 requirements. Specifically, the chance constraint $\mathbb{P}(\cdot)\geq 1-\varepsilon$ ensures that the probability of jointly satisfying these constraints within a given hour is at least $1-\varepsilon$. In accordance with the P90 requirement, $\varepsilon=0.1$. The bids are submitted on an hourly basis (though in Denmark this will be changed to 15-minute intervals), and the proposed model accounts only for the reservation of flexibility capacity. Consequently, the rebound effect is not considered, implying no inter-temporal dependency within the hour in the bidding scheme. Future work could address this limitation by incorporating activation data.

Chance-constrained programs are generally computationally intractable. However, if the underlying probability distribution satisfies certain properties, the chance constraints can be reformulated analytically. For instance, methods such as the Bonferroni approximation \citep{BonferroniCorr} replace probabilistic constraints with a series of more tractable ones, simplifying the problem by evaluating only the marginal distributions of the random variables, albeit at the cost of precision. This means that any correlation structure among the random variables $\{R^{\uparrow}, R^{\downarrow}, R^{\rm{E_{20}}}\}$ in the joint chance constraint is disregarded. Alternatively, sample-based methods capture uncertainty through a set of representative scenarios, enabling optimization that explicitly accounts for a range of possible outcomes. Both approaches allow EV aggregators to design bidding strategies that manage the stochastic nature of their resources while respecting the operational requirements of ancillary service markets.

\begin{figure*}[t]
    \centering
    \includegraphics[width=0.53\textwidth]{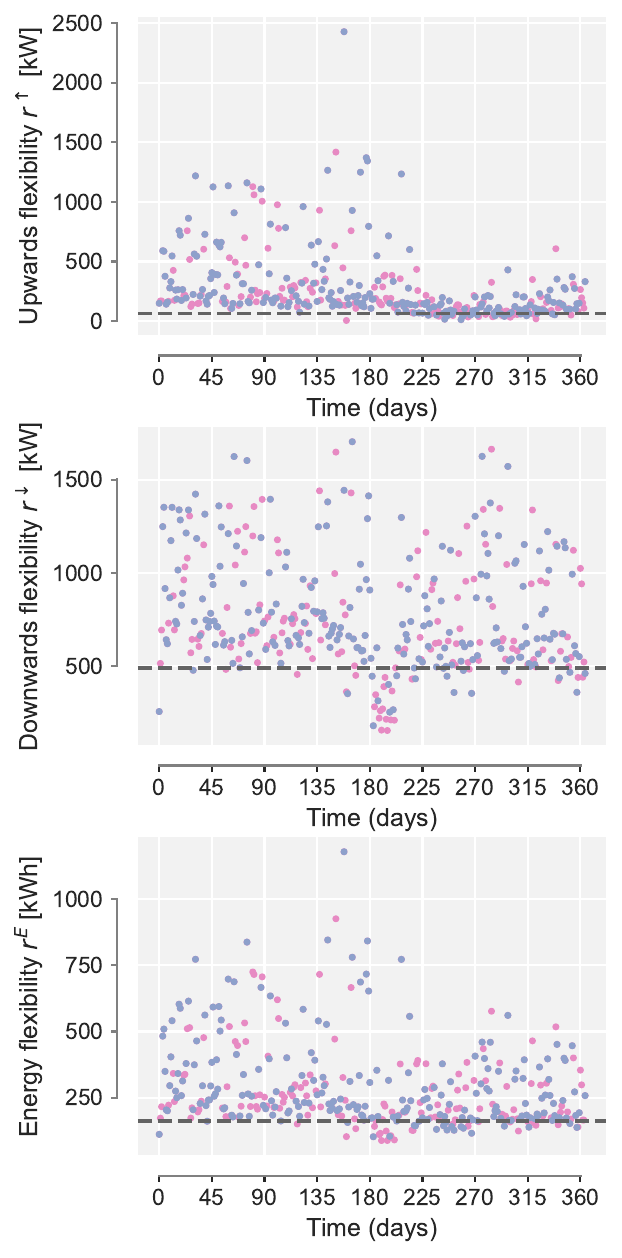}
    \caption{Scatter plots illustrating the estimated upward (top), downward (middle), and energy (bottom) flexibilities for a representative hour across all days of the year. Blue samples represent the in-sample data used for optimization in a given run, while pink samples correspond to the ex-post out-of-sample validation in the same run. The dashed line marks the empirical 10\textsuperscript{th} percentile of the blue dots. Consistent with extreme value theory, only the blue samples below this percentile are used for estimation in the analytical method.}
    \label{fig:scatter time}
\end{figure*}

Figure \ref{fig:scatter time} presents scatter plots of the in-sample and out-of-sample data used in our case study. The in-sample data (blue samples) are used in the optimization problem to determine the optimal bids, while the out-of-sample data (pink samples) are used ex-post to evaluate these bids. Extreme value theory \citep{EVTCastillo} studies the behavior of extreme values (maxima or minima) in a dataset, particularly as the sample size increases. In this study, the extreme values of interest correspond to the blue samples below the dashed line, which represents the empirical 10\textsuperscript{th} percentile of the in-sample data. To reduce bias, the model is solved 10 times, each with a different set of in-sample and out-of-sample data, while keeping the number of samples in each set constant.

To derive an analytical reformulation of the optimization problem \eqref{eq:MAIN opt form}, we fit a distribution to the tail data (blue samples below the empirical 10\textsuperscript{th} percentile) that enables an analytical representation of the chance constraint. Among the options considered for modeling the tail, we tested both the Weibull and Generalized Pareto distributions. The Weibull distribution was preferred because it is comparatively more straightforward to fit and tune, yields a lower negative log-likelihood when fitted via maximum likelihood, and demonstrates better out-of-sample performance. Consequently, our analysis is based on the Weibull distribution.


It is worth noting that in the extreme value theory literature, the generalized extreme value distribution is more commonly used, as it arises as the limiting distribution of the maximum of $n$ independent and identically distributed random variables as $n$ becomes large. However, convergence to this limiting distribution can be slow. Moreover, the generalized extreme value distribution introduces additional parameters that must be estimated, often requiring a larger number of samples.



\section{Distribution fitting and validation}\label{sec:analyticaltheory}

This section outlines the methodology for fitting a probability distribution to our data, providing the necessary input for the analytical solution presented in Section \ref{sec:analyticalsolution}.

\subsection{Fitting a Weibull distribution to the tail data}\label{sec:fitting weibull}
Aiming to obtain an analytical solution, we fit a Weibull distribution to the tails of our data. The rationale for choosing this distribution lies in its flexible shape parameters, allowing to capture a wide range of tail behaviors. This flexibility is particularly advantageous, as this distribution can effectively model both skewed and heavy-tailed data. Its ability to adapt to both light and heavy tails enables us to accurately capture the realistic variability and potential extremes in the data. The Weibull distribution has extensive applications in fields such as reliability engineering, where it models failure times and lifetimes, and in risk management for assessing minimum extreme events. 

\begin{figure}
    \centering
    \includegraphics[width=0.8\linewidth]{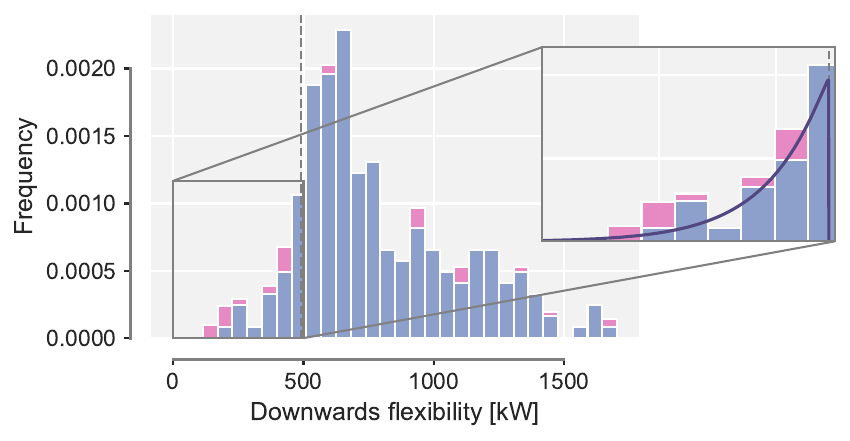}
    \caption{Distribution of the downwards flexibility at the arbitrarily selected hour 13. The blue distribution is constructed using in-sample data, while the pink part represents the out-of-sample data. A Weibull distribution with the shape parameter $\hat{\gamma}=0.9265$ is fitted to the lower tail of the blue distribution. The vertical dashed line indicates the empirical 10\textsuperscript{th} percentile value of the in-sample data, below which the Weibull distribution is fitted.}
    \label{fig:FittedWeibullHour13}
\end{figure}

For each flexibility $\{R^\uparrow, R^\downarrow, R^{\rm{E_{20}}}\}$, we select a subset of realizations for each hour for which we fit a distribution to the tail. As illustrated in Figure \ref{fig:scatter time}, the relevant tail, in line with the case study and the P90 requirement, consists of all in-sample data points below the 10\textsuperscript{th} percentile, $r \leq r_{\varepsilon}=r_{0.1}$, where we let $r$ denote a realization of an arbitrary flexibility. Since the data in this tail does not resemble a Weibull distribution, we transform the data using $x=-r + r_{0.1}$, to which the Weibull distribution is more appropriately fitted. Figure \ref{fig:FittedWeibullHour13} shows the distribution of realizations of the estimated downward flexibility, $r^\downarrow$, for a randomly selected hour. In line with Figure \ref{fig:scatter time}, the blue bars represent the distribution based on in-sample data, while the pink bars correspond to the remaining out-of-sample data. We aim to fit a distribution based on the in-sample data. The threshold of the in-sample distribution, below which the Weibull distribution is fitted, i.e., the 10\textsuperscript{th} percentile of the sampled realizations, is indicated by a vertical dashed line.

We begin the procedure of fitting the Weibull distribution \citep{MLE} by defining its two-parameter probability density function (PDF) as follows:
\begingroup
\allowdisplaybreaks
\begin{subequations}
  \begin{align}
   f(x)&=\begin{cases}\kappa\gamma x^{\gamma-1}\exp(-\kappa x^{\gamma}),&x\geq 0,\\0,&x<0,
    \end{cases} \label{eq:WeibullPDF}\\ 
\intertext{where the scale and shape parameters, $\kappa, \gamma>0$, can be estimated by maximum likelihood techniques. The likelihood function of \eqref{eq:WeibullPDF} is defined for $x\geq 0$ as:}
   &\mathcal{L}(x;\kappa,\gamma)=\prod_{i=1}^n \kappa\gamma x_i^{\gamma-1}\exp(-\kappa x_i^\gamma), \\
\intertext{where $n$ is the number of samples used for the estimation. Taking the logarithm of the likelihood function, then differentiating with respect to $\kappa$ and $\gamma$ in turn, and setting the derivatives equal to zero, we obtain the maximum likelihood estimator for $\kappa$ in terms of the estimator of $\gamma$ as:}
    &\quad\quad\hat{\kappa}(\hat{\gamma})=\frac{n}{\sum_{i=1}^n x_i^{\hat{\gamma}}},\label{eq:WeibullKappaEst} \\
\intertext{where every symbol with a hat denotes the estimator of the corresponding variable. Substituting \eqref{eq:WeibullKappaEst} into the log-likelihood function, we obtain:}
    \ell_n(x; \hat{\kappa}(\hat{\gamma}), \hat{\gamma})&=n\left(\log n - \log\left(\sum_i x_i^{\hat{\gamma}}\right) + \log\hat{\gamma} - 1\right)+  \nonumber \\&\quad\quad\quad\quad(\hat{\gamma}-1)\sum_i \log x_i^{\hat{\gamma}}, \label{eq:loglikfun} 
  \end{align}
\end{subequations}
which we maximize over a range of appropriate $\gamma$-values. Looking back at Figure \ref{fig:FittedWeibullHour13}, the Weibull distribution is fitted to the lower tail of the sampled data, with the shape parameter $\hat{\gamma}=0.9265$.
As it is needed in later analysis, we define the cumulative distribution function (CDF) of the Weibull distribution as: 
\begin{subequations}
    \begin{align}
        F(x)&=\begin{cases}1-\exp(-\kappa x^{\gamma}),&x\geq 0,\\0,&x<0.
    \end{cases} \label{eq:WeibullCDF} \\
\intertext{We also define the tail distribution function, otherwise known as the survival function:}
        \Bar{F}(x)=1-&F(x)=\begin{cases}\exp(-\kappa x^{\gamma}),&x\geq 0,\\0,&x<0.
    \end{cases} \label{eq:WeibullCDFtail}
    \end{align}
\end{subequations}
\endgroup

\subsection{Validation of distribution fits}\label{sec:distribution validation}
To validate whether the fitted distribution correctly reflects the data, we use the Kolmogorov-Smirnov (KS) test \citep{KSTest}. This is a nonparametric test used to compare the equality of continuous, one-dimensional probability distributions, and it tests whether a sample comes from a specified reference distribution. The test statistic reports the maximum absolute difference between the empirical CDF, $F_{n}(x)$, and the reference CDF, $F(x)$, as:
\begin{equation}\label{eq:KStest}
     D_{n}=\sup _{x}|F_{n}(x)-F(x)|,
\end{equation}
where the empirical CDF is defined as:
\begin{equation}
    F_{n}(x)={\frac{1}{n}}\sum _{i=1}^{n}\mathbf {1} _{X_{i}\leq x}.
\end{equation}

As before, $n$ denotes the number of samples used in the estimation. Using \eqref{eq:KStest}, we can determine whether to reject or accept the hypothesis that the realizations originate from the reference distribution. In this context, the null hypothesis $H_0$ assumes that the data follow the specified reference distribution, whereas the alternative hypothesis $H_1$ assumes that they do not.

The KS test calculates the $D_n$ statistic and compares it to a critical value or uses it to compute a $p$-value based on the sample size and the reference distribution. If the $D_n$ statistic exceeds the critical value or the $p$-value falls below a pre-specified significance level (e.g., 0.05), $H_0$ is rejected, suggesting that the data is unlikely to come from the reference distribution. Otherwise, there is insufficient evidence to reject $H_0$, meaning the data is consistent with the specified distribution. This makes the KS test a robust tool for validating the goodness-of-fit between a dataset and a proposed model.

To further compare the goodness-of-fit between several distributions, such as the fit of the Weibull distribution for a range of parameter values, the negative log-likelihood (NLL) value is utilized, where a lower value signifies a better fit. This is because the NLL effectively quantifies how well a given model and its parameters explain the observed data, with lower NLL values corresponding to higher likelihoods of the observed data under the model. In essence, we use the negative of \eqref{eq:loglikfun} and compare the resulting value from a range of $\gamma$-values, and choose the $\gamma$ which gives us the lowest \citep{MLNLL}. By using the NLL and the KS test, we validate that the obtained distribution fits are satisfying for our analysis.

\subsection{Validation results} \label{sec:results from KS and emp perc}
\begin{figure}
    \centering
    \includegraphics[width=\linewidth]{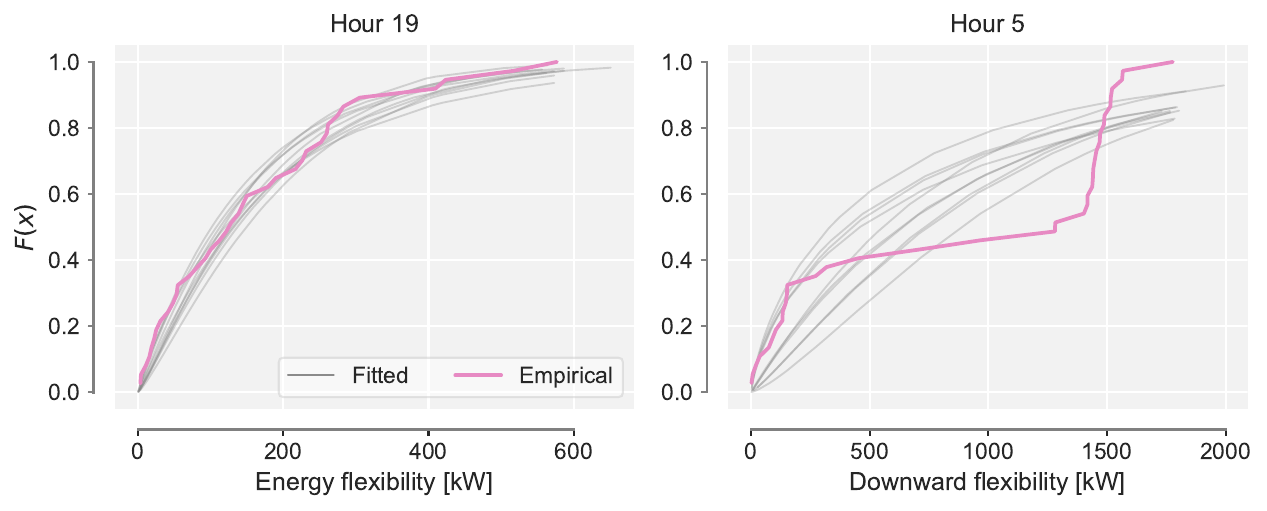}
    \caption{Best (left) and worst (right) distribution fits (grey) compared to all data below the 10th percentile (pink). The $p$-values for the best fit range from 0.871 to 0.999, while for the worst fit they range from 0.218 to 0.871.}
    \label{fig:CDF_results}
\end{figure}

Figure \ref{fig:CDF_results} presents the results of the distribution fits for the best and worst average fits over 10 runs, based on the validation metrics described in Section \ref{sec:distribution validation}. Recall that each run uses a different set of in-sample data, while keeping the number of samples constant. Additionally, average results across all hours for all flexibilities are provided in Table \ref{tab:D and p-val} in Appendix B. The $p$-values associated with the maximum absolute difference ($D_n$) indicate that fits from all runs are accepted ($p > 0.05$), albeit with varying strengths. For the best fit, $p$-values range from 0.871 to 0.999, with an average maximum absolute difference of 0.105, whereas for the worst fit, $p$-values range from 0.218 to 0.871, with an average maximum absolute difference of 0.273.

To test the accuracy of the empirical percentile $r_{0.1}$, we drew 216 samples without replacement from the data pool and repeated this procedure 5000 times. The choice of 216 samples will be justified later in Section~\ref{sec:samplingmethod}. For each run, we computed the empirical percentile, then obtained the mean and standard deviation across all runs. From these, we calculated the coefficient of variation, which measures variation relative to the mean. The relative variation lies between $1.4\%$ and $10.9\%$, with the maximum observed for upward flexibility at hour~10 and the minimum for energy flexibility at hour 19, which is also the case where the Weibull distribution provides the best fit.

\section{Analytical solution}\label{sec:analyticalsolution}

This section presents the proposed analytical solution for addressing the chance-constrained bidding problem introduced in Section \ref{sec:m&m}.

\subsection{Bonferroni correction}
The joint chance constraint in \eqref{eq:MAIN opt form} can be decomposed into three individual constraints using the Bonferroni correction \citep{BonferroniCorr}. This adjustment redistributes the overall allowance for violation equally across each constraint, effectively lowering the significance level for each test to ensure that the total probability of violation remains within the desired limit:
\begin{subequations}\label{eq:BonferroniCorrectedConstraints}
    \begin{align}
        \mathbb{P}&\left(R^\uparrow \leq 0.2b^\downarrow + b^\uparrow\right) \leq \frac{\varepsilon}{3}, \\
        \mathbb{P}&\left(R^\downarrow \leq b^\downarrow\right) \leq \frac{\varepsilon}{3}, \\
        \mathbb{P}&\left(R^{\rm{E_{20}}}\leq b^\downarrow \right) \leq \frac{\varepsilon}{3}.
    \end{align}
\end{subequations}



We conceptualize the method using hypothesis testing. The Bonferroni correction establishes an upper bound on the overall risk of incorrectly rejecting at least one hypothesis from the entire set being tested. The bound is valid regardless of any relationships between the hypotheses, but it is often larger than the true error probability. Specifically, for testing three hypotheses at a significance level $\varepsilon$, each hypothesis is rejected if its individual error probability is less than $\varepsilon/3$. The technique also accounts for data dependencies, despite its inherently conservative nature.

The Bonferroni approach essentially replaces the solution space for each decision variable in \eqref{eq:CC2} with a smaller set, as presented in \eqref{eq:BonferroniCorrectedConstraints}, which may make the solution more conservative. On the other hand, one might replace \eqref{eq:CC2} with a larger solution space by setting the right-hand side of the inequalities in \eqref{eq:BonferroniCorrectedConstraints} to $\varepsilon$. These sets may act as lower and upper bounds on the solution of \eqref{eq:MAIN opt form}, respectively. We elaborate with a two-dimensional example: using the joint probability $\mathbb{P}(x \leq X, y \leq Y) \geq 1 - \varepsilon$, shaded in gray in Figure \ref{fig:2d illustrative example}, we can restate it as:
\begingroup
\allowdisplaybreaks
\begin{subequations}
    \begin{align}
  \mathbb{P}\left(X < x\right) + \mathbb{P}\left(Y < y\right) &- \mathbb{P}\left(X<x, Y<y\right) < \varepsilon \label{eq:P10} \\
\intertext{which is the area shaded in pink and purple, or,}
    \mathbb{P}\left(X < x\right) + \mathbb{P}\left(Y < y\right) &< \varepsilon + \mathbb{P}\left(X<x, Y<y\right).  \label{eq:P10 2} \\
\intertext{We enforce the stricter inequality}
    \mathbb{P}\left(X < x\right) + &\mathbb{P}\left(Y < y\right) < \varepsilon, \label{eq:P10 strickter}\\
\intertext{which implies that the probability mass in the joint area, $\mathbb{P}\left(X<x, Y<y\right)$, is zero, and apply the Bonferroni correction as:}
    \begin{split} \label{eq:bonferronigeneric case}
        \mathbb{P}\left(X < x\right)& < \frac{\varepsilon}{2} \\
        \mathbb{P}\left(Y < y\right)& < \frac{\varepsilon}{2}.
    \end{split}
    \end{align}
\end{subequations}
\endgroup

Any solution of $x$ and $y$ which satisfies \eqref{eq:bonferronigeneric case} will then satisfy \eqref{eq:P10 strickter} $\Rightarrow$ \eqref{eq:P10 2} $\Rightarrow$ \eqref{eq:P10}. The inequality \eqref{eq:P10 strickter} implies that threshold values of failure probability are distributed over any combination of $x$ and $y$, meaning that $x$ and $y$ may take any values satisfying $\mathbb{P}(X\leq x) < \epsilon$ or $\mathbb{P}(Y\leq y)< \epsilon$ as long as the sum of the two probabilities also satisfies \eqref{eq:P10 strickter}. The Bonferroni corrected version \eqref{eq:bonferronigeneric case} expresses that at most $\varepsilon/2$ of the failure can be associated with each variable, meaning that the possible values of $x$ and $y$ have a tighter constraint. If the distribution of $X$ and $Y$ are the same, while the intersect is zero, then this is the exact solution where \eqref{eq:bonferronigeneric case} = \eqref{eq:P10 strickter}. In Figure \ref{fig:2d illustrative example}, this is the case if we assume that the probability mass in the pink and purple shaded areas (excluding the joint probability area) are equal, i.e., $\mathbb{P}(X\leq x) = \mathbb{P}(Y\leq y)$ and $\mathbb{P}\left(X<x, Y<y\right)=0$.

\begin{figure}[t]
  \centering
  \begin{tikzpicture}
  \pgfdeclarepatternformonly{north east lines wide}%
   {\pgfqpoint{-1pt}{-1pt}}%
   {\pgfqpoint{10pt}{10pt}}%
   {\pgfqpoint{5pt}{5pt}}%
   {
     \pgfsetlinewidth{0.6pt}
     \pgfpathmoveto{\pgfqpoint{0pt}{0pt}}
     \pgfpathlineto{\pgfqpoint{9.1pt}{9.1pt}}
     \pgfusepath{stroke}
    }

    \pgfdeclarepatternformonly{north west lines wide}%
   {\pgfqpoint{-1pt}{-1pt}}%
   {\pgfqpoint{10pt}{10pt}}%
   {\pgfqpoint{5pt}{5pt}}%
   {
     \pgfsetlinewidth{0.6pt}
     \pgfpathmoveto{\pgfqpoint{0pt}{9.1pt}}
     \pgfpathlineto{\pgfqpoint{9.1pt}{0pt}}
     \pgfusepath{stroke}
    }
    \begin{axis}[
        name=myAxis,
        xmin=0, xmax=3,
        ymin=0, ymax=3,
        ytick=\empty,
        xtick={0.5}, ytick={0.5},
        xticklabels={$x$}, yticklabels={$y$},
      ]  
      \addplot[black, samples=100, domain=0:3, name path=A] {0.5}; 
      \addplot[samples=50, domain=0:3, name path=B] {3}; 
      \path[name path=xaxis] (\pgfkeysvalueof{/pgfplots/xmin}, 0) -- (\pgfkeysvalueof{/pgfplots/xmax},0);
       \addplot[pattern color=red!30!white, pattern=north west lines wide] fill between[of=B and xaxis, soft clip={domain=0:0.5}];
       \addplot[pattern color=pink, fill=gray!20] fill between[of=B and A, soft clip={domain=0.5:3}];
      \addplot[pattern color=violet!30!white, pattern=north east lines wide] fill between[of=A and xaxis, soft clip={domain=0:3}];
      \node[text width=3cm] at (1.75,1.75) {$\mathbb{P}(X>x, Y>y)$};
      \node[text width=2cm] at (1.75,0.25) {$\mathbb{P}(Y\leq y)$};
      \node[text width=2cm, rotate=90] at (0.25,1.75) {$\mathbb{P}(X\leq x)$};
      \addplot [black,mark=none] coordinates {(0.5, -1) (0.5, 3)};
      \end{axis}
      \draw[->] ([yshift=-1ex] myAxis.south west) -- (0.55,0.5);
      \node[left, align=center, text=black]
        at ([yshift=-2.5ex] myAxis.south west) {$\mathbb{P}(X\leq x, Y\leq y)$};
  \end{tikzpicture}
  \caption{A simple two-dimensional illustrative example of a joint distribution.} \label{fig:2d illustrative example}
\end{figure}

\subsection{Analytical reformulation of \eqref{eq:MAIN opt form}}
Following the procedure in Section \ref{sec:analyticaltheory}, the Weibull distribution is fitted to data under a threshold level, $r_{0.1}$, which in the present case is the empirical 10\textsuperscript{th} percentile of each flexibility, i.e., $\mathbb{P}(R \leq r_{0.1})=\varepsilon$. Applying Bayes' theorem, we can thus express the conditional probability for an arbitrary random variable $R$ being under a certain threshold $b$ as:
\begin{equation}
    \begin{split}
        &\mathbb{P}\left(R \leq b \mid R \leq r_{0.1}\right) = \mathbb{P}\left(R \leq b\right)/\mathbb{P}\left(R \leq r_{0.1}\right) \\
        \Rightarrow& \mathbb{P}\left(R \leq b\right) = \mathbb{P}\left(R \leq r_{0.1}\right)\mathbb{P}\left(R \leq b \mid R \leq r_{0.1}\right),
    \end{split}
\end{equation}
where it is implicit that $\mathbb{P}\left(R \leq r_{0.1} \mid R \leq b\right) = 1$. This gives a general formulation of the corrected constraints  \eqref{eq:BonferroniCorrectedConstraints} as:
\begin{equation}\label{eq:Bonferroni distributed constraints}
    \varepsilon \mathbb{P}\left(R \leq b \mid R \leq r_{0.1}\right) \leq \frac{\varepsilon}{3},
\end{equation}
which, after inserting the tail of the CDF in \eqref{eq:WeibullCDFtail}, simplifies to:
\begin{equation}
    b\leq r_{0.1} -\left(\frac{1}{\kappa}\log3 \right)^{1/\gamma}.
\end{equation}

\begin{figure}
    \centering
    \includegraphics[width=0.66\linewidth]{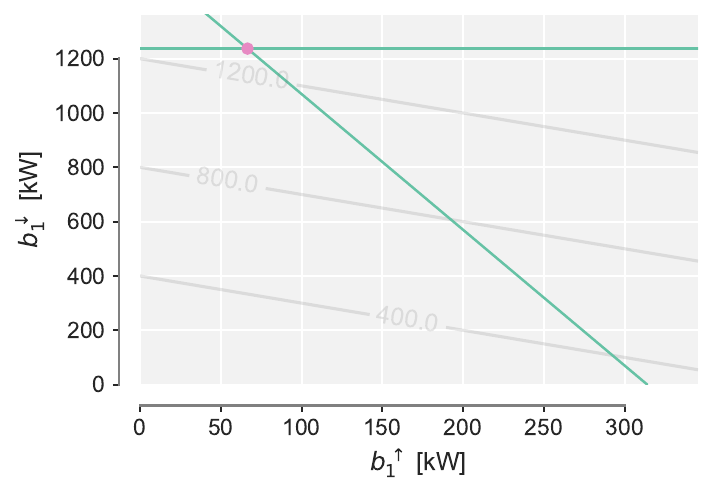}
    \caption{Geometric illustration of the optimization problem \eqref{eq:bonferroni opt prob} for the arbitrarily selected hour 1. The grey lines are the contours of the objective function, and the green lines are the constraints, where the two constraints on $b^\downarrow$, \eqref{eq:bonferroni opt prob CC2} and \eqref{eq:bonferroni opt prob CC3}, are combined into a single constraint as the minimum of the two, i.e., $b^{\downarrow} \leq \min\{\eqref{eq:bonferroni opt prob CC2}, \eqref{eq:bonferroni opt prob CC3}\}$. The point where the green lines intersect indicates the optimal solution.}
    \label{fig:geometric_solution}
\end{figure}

\begingroup
\allowdisplaybreaks
Using this formulation, with separately fitted distributions for each of the three flexibilities following the procedure outlined above, and applying the Bonferroni correction, the optimization problem  \eqref{eq:MAIN opt form} can be rewritten  as:
\begin{subequations}\label{eq:bonferroni opt prob}
    \begin{align}
    &\max_{b^\downarrow, b^\uparrow \geq 0} b^\uparrow + b^\downarrow \\
        \text{s.t.}\quad& \nonumber\\
         &0.2b^{\downarrow} + b^{\uparrow} \leq r_{0.1}^\uparrow -\left(\frac{1}{\kappa^\uparrow}\log 3 \right)^{1/\gamma^\uparrow} \label{eq:bonferroni opt prob CC1}\\
        &b^{\downarrow} \leq r_{0.1}^\downarrow -\left(\frac{1}{\kappa^\downarrow}\log 3\right)^{1/\gamma^\downarrow} \label{eq:bonferroni opt prob CC2} \\
        &b^{\downarrow} \leq r_{0.1}^{\rm{E_{20}}} -\left(\frac{1}{\kappa^{\rm{E}}}\log 3\right)^{1/\gamma^{\rm{E}}}, \label{eq:bonferroni opt prob CC3}
    \end{align} 
\end{subequations}
where the two latter constraints can be combined as $b^{\downarrow} \leq \min\{\eqref{eq:bonferroni opt prob CC2}, \eqref{eq:bonferroni opt prob CC3}\}$. Using the correction method, the feasible region of the original problem  \eqref{eq:MAIN opt form} is reduced, meaning that any solution satisfying \eqref{eq:bonferroni opt prob} will also satisfy \eqref{eq:MAIN opt form}. The optimization problem \eqref{eq:bonferroni opt prob} is two-dimensional with linear constraints, and can be easily solved analytically. Figure \ref{fig:geometric_solution} provides a graphical representation of the optimization problem for the arbitrarily selected hour 1, illustrating how the optimal solution is found geometrically.
\endgroup

\section{Sample-based solution for benchmarking}\label{sec:SAAform}
As a benchmark for our proposed analytical approach, we use the commonly adopted sample-based solution to reformulate \eqref{eq:MAIN opt form}.

\subsection{Sample-based reformulation of the optimization problem}
\begingroup
\allowdisplaybreaks

We reformulate the joint chance-constrained optimization problem  \eqref{eq:MAIN opt form} using $|\Omega|$ samples, where each sample $i$ includes the realizations of the three random variables $\{r_i^\uparrow, r_i^\downarrow, r_i^{\rm{E_{20}}}\}$. This reformulation is written as: 
\begin{subequations}
\begin{align}
        &\max_{b^\downarrow, b^\uparrow \geq 0} b^\uparrow + b^\downarrow \\
        \text{s.t.}\quad& \nonumber\\
        &0.2b^\downarrow + b^\uparrow - r^{\uparrow}_i \leq M^\uparrow y_i,\quad i=1,\dots,|\Omega| \label{eq:MILPcons1} \\
        &b^\downarrow - r^{\downarrow}_i \leq M^\downarrow y_i,\quad i=1,\dots,|\Omega| \label{eq:MILPcons2} \\
        &b^\downarrow - r^{\rm{E_{20}}}_i \leq M^{\rm{E_{20}}} y_i,\quad i=1,\dots,|\Omega| \label{eq:MILPcons3} \\
        &\sum_{i=1}^{|\Omega|} y_i \leq |\Omega|\varepsilon, \label{eq:MILPcons4} \\
        &y_i \in \{0,1\}, \quad i=1,\dots,|\Omega|,\label{eq:MILPcons5}
\end{align}
\end{subequations}
where $\{M^\uparrow, M^\downarrow, M^{\rm{E_{20}}}\}$ are sufficiently large positive constants, and $y_i$ is an auxiliary binary variable corresponding to sample $i$. Constraints \eqref{eq:MILPcons1}-\eqref{eq:MILPcons3} ensure that if $y_i=0$, all constraints are fulfilled for a specific sample $i$, while if $y_i=1$, these constraints are relaxed.  Constraint \eqref{eq:MILPcons4} ensures that the total budget for constraint violations is preserved, allowing for at most a 10\% violation of the samples, in accordance with the P90 requirement. Due to the reduced number of data points in the case study, the problem remains tractable and can be solved without the need for additional simplification or relaxation.
\endgroup

\subsection{Sampling}\label{sec:samplingmethod}
To determine the minimum number of randomly selected samples necessary to accurately represent the estimated data, the following formula is used \citep{SampleSizeSAA}: 
\begin{equation}\label{eq:samplesize}
    |\Omega|\geq \frac{2}{\varepsilon}\log(\delta^{-1})+2p + \frac{2p}{\varepsilon}\log(2/\varepsilon),
\end{equation}
where $p$ is the number of decision variables in the optimization problem (two in our case), $\varepsilon = 0.1$ as per the P90 requirement, and by setting $\delta=0.01$, we achieve a 99\% confidence level that the set of samples accurately represents the underlying distribution of the data. This results in $|\Omega|=216$ samples (approximately 60\% of all realizations) being used for the optimization of bids across each hour, with the remaining 150 samples (approximately 40\%) reserved for out-of-sample validation, as described in the following subsection.

\subsection{Ex-post out-of-sample validation of constraints}\label{sec:OOSvalidation}
For the validation of the optimal reserve capacity bids obtained, i.e., $b^{\uparrow*}$ and $b^{\downarrow*}$, we construct a sample set  $\Phi\notin\Omega$ and quantify the out-of-sample constraint violations. To do this, we count the number of instances where one or multiple of the following inequalities are violated for the realization of $\{r^\uparrow_{i}, r^\downarrow_{i}$, $r^{\rm{E_{20}}}_{i}\}$, where $i=1,\dots,|\Phi|$:
\begin{subequations}\label{eq:OOSvalidation}
    \begin{align}
        0.2b^{\downarrow*} + b^{\uparrow*} - r^\uparrow_{i} > 0, \\
        b^{\downarrow*} - r^\downarrow_{i} > 0, \\
        b^{\downarrow*} - r^{\rm{E_{20}}}_{i} > 0.
    \end{align}
\end{subequations}

Any violation of \eqref{eq:OOSvalidation} for a given realization is counted as a violation of the joint chance constraint \eqref{eq:CC2}. If more than one of the inequalities is violated simultaneously, it is counted as a single violation.

As a summary, the sampling, distribution fitting, decision making, and out-of-sample validation procedure is illustrated in Figure \ref{fig:flowchart sampling}.
The randomly sampled data points are used in both the analytical and sample-based approaches, with the former utilizing only a selected subset of the data points, as detailed in Section \ref{sec:analyticaltheory}.

\begin{figure}[t]
  \begin{center}
    \begin{tikzpicture}[
    node distance = 8mm and 12mm,
         N/.style = {draw, fill=white, minimum size=1.2cm, align=center},
        every edge/.append style = {draw, semithick, -Stealth}
                            ]
    \node[N] (all) {All 366 samples};
    \node[N, right=of all, two copy shadow] (sample) {216 random\\samples};
    \node[N, right=of sample, two copy shadow] (percentile) {Samples under \\10th percentile};
    \node[N, right=of percentile] (weibull) {Distribution\\fitting};
    \node[N, below=of weibull] (parameters) {In-sample\\optimization\\problem \eqref{eq:bonferroni opt prob}};
    \node[N, below=of sample] (validation) {OOS bid\\violations \eqref{eq:OOSvalidation}};
    \node[N, below=of all, two copy shadow] (oos) {150 remaining\\samples};
    
    \node[N, below=of percentile] (KS) {KS test \eqref{eq:KStest}};
    \draw[->]  (all.east) -- (sample.west);
    \draw[->]  (all.south) -- (oos.north);
    \draw[->]  (sample.east) -- (percentile.west);
    \draw[->]  (sample.south) -- (validation.north);
    \draw[->]  (percentile.east) -- (weibull.west);
    \draw[->]  (weibull.south west) -- (KS.north east);
    \draw[->]  (weibull.south) -- node [midway, left, xshift=-10pt] {$\hat{\kappa}, \hat{\gamma}$} (parameters.north);
    \draw[->]  (percentile.south) -- (KS.north);
    \draw[->]  (oos.east) -- (validation.west);
    \draw[->]  (parameters.south) -- +(0,-0.5) -| (validation.south);
    \path (parameters) -- (validation) coordinate[midway] (midpoint);
    \node[below=18pt of midpoint] {$b^{\uparrow *}, b^{\downarrow *}$};

      \begin{scope}[on background layer]
    \node[fill=red!20,inner sep=10pt,rounded corners=0.5cm,fit=(sample) (validation)] {};
    \node[fill=red!20,inner sep=10pt, inner ysep=16pt,rounded corners=0.5cm,fit=(validation) (oos)] (red) {};
    \node[fill=blue!20,inner sep=10pt,rounded corners=0.5cm,fit=(percentile) (weibull) (parameters) (KS)] (blue) {};
    \node[below=of red, yshift=.5cm, inner sep=0] {\textcolor{red}{Sampling and OOS validation}}; 
  \end{scope}
      \node[below=of blue, yshift=.5cm, inner sep=0, text width=5cm, align=center] {\textcolor{blue}{Distribution fitting and \\ in-sample optimization problem}}; 
    \end{tikzpicture}
    \end{center}
  \caption{Flowchart of the sampling, distribution fitting, decision making, and out-of-sample (OOS) validation procedure. From the 366 available data points, 216 are randomly sampled without replacement. Among these, the 22 samples below the empirical 10th percentile are used to fit a Weibull distribution via maximum likelihood estimation. The estimated parameters, $\hat{\kappa}$ and $\hat{\gamma}$, are then assessed with a KS test against the same 22 samples. The remaining 150 samples are employed for out-of-sample analysis of bid violations under both the sample-based and analytical optimization approaches.}
  \label{fig:flowchart sampling}
\end{figure}

\section{Numerical results}\label{sec:results}

Our case study relies on real-world data comprising charging profiles from 1400 residential EV charging boxes in Denmark. The measurements were collected between March 24, 2022, and March 21, 2023, with an average time step of 2.84 minutes. To ensure a uniform resolution, the data were interpolated to one-minute granularity. It is assumed that each charging box corresponds to a single EV. The historical EV consumption levels serve as the baseline for estimating flexibility. Further details of the case study are provided in Appendix~\ref{sec:case study intro}.

In this section, we first present numerical results regarding the bidding strategy of the EV aggregator, obtained from both the analytical and sample-based approaches. We also report the rate of out-of-sample constraint violations and provide a sensitivity analysis with respect to $\varepsilon$. 
All source codes are publicly available in \citet{Git}, while the historical EV consumption data remain proprietary. The estimated values of $\hat{\kappa}$ and $\hat{\gamma}$ from the distribution fittings are also reported in \citet{Git}.

All optimization was performed on an HP EliteBook 840 14-inch G10 Notebook PC, equipped with a 13th Gen Intel\textsuperscript{\textregistered} Core\texttrademark~i7-1365U processor (10 cores, 12 logical processors), using Visual Studio Code. The sample-based optimization completes in 44.09~seconds, averaging 1.84~seconds per hour (10 runs with different sample splits). In comparison, the optimization based on the analytical formulation of the joint chance constraint runs in 9.19~seconds, averaging 0.38~seconds per hour over the same 10 runs, corresponding to a speed-up factor of 4.8 relative to the sample-based approach. When including the time required for distribution fitting, the full analytical procedure completes in 32.9 seconds, averaging 1.33~seconds per hour, which corresponds to a speed-up factor of 1.34--1.38 compared to the sample-based method.

\begin{figure*}[t]
    \centering
    \includegraphics[width=\textwidth]{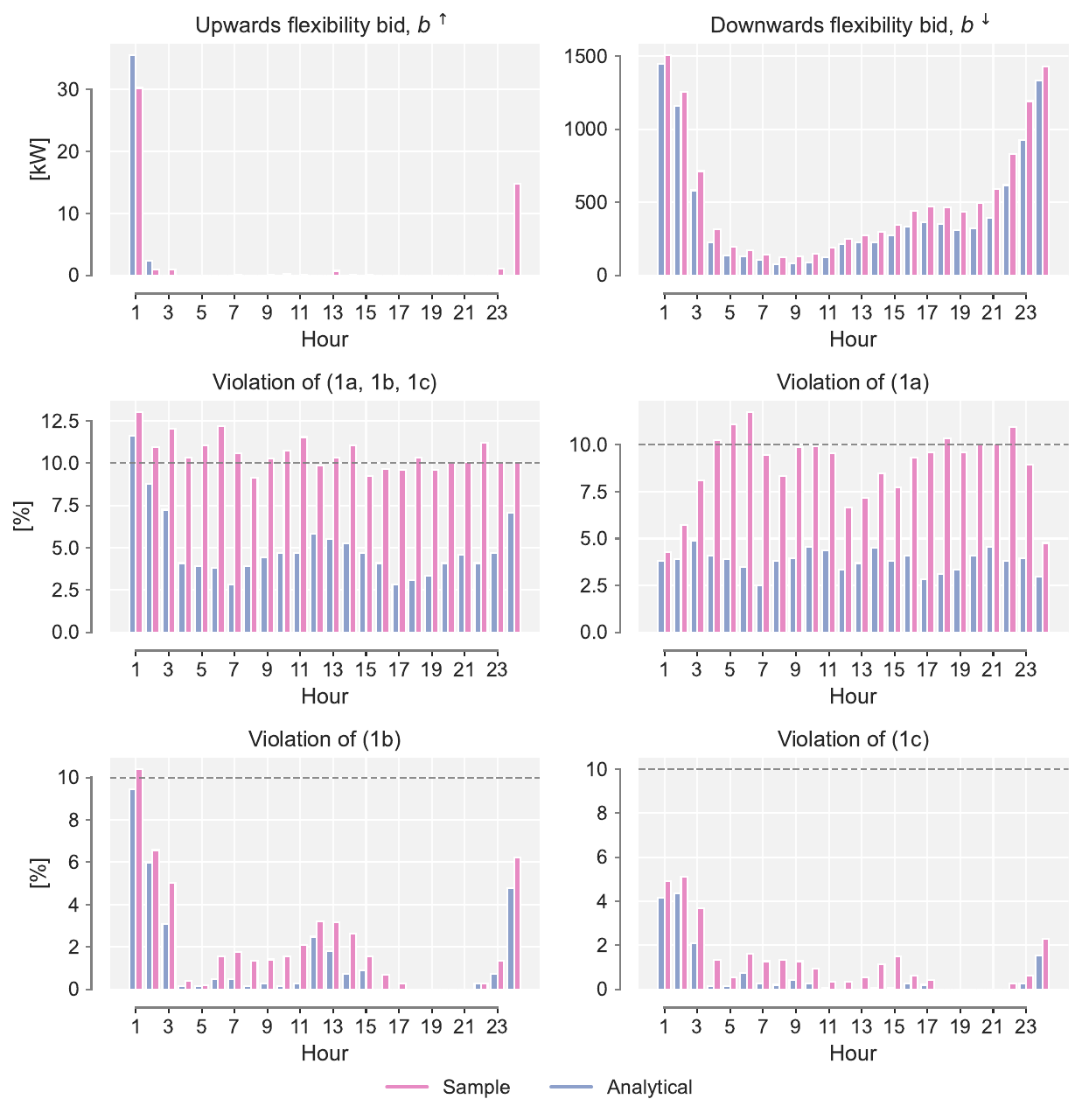}
    \caption{The two top plots illustrate the capacity bids for FCR-D up (left) and FCR-D down (right) services obtained with the proposed analytical method (blue) and the benchmark sample-based method (pink). The remaining four plots show the out-of-sample constraint violation rates, reported separately for each constraint as well as in aggregate. All results represent the mean over 10 runs, where each run uses different in-sample and out-of-sample sets, while the number of samples in each set remains fixed.}
    \label{fig:results_all}
\end{figure*}

\subsection{Bidding strategy}
As described earlier, we consider 10 runs, each based on 216 randomly selected in-sample data points drawn from the 366 available points. The top plots of Figure \ref{fig:results_all} illustrate the average reserve capacity bids across all runs for the FCR-D up and down markets. The remaining plots report the average constraint violation rates, obtained through the out-of-sample validation procedure outlined in Section \ref{sec:OOSvalidation}, using 150 samples not included in the optimization.

As expected from the validation discussion in Section~\ref{sec:results from KS and emp perc}, the variation of the empirical percentile does not affect the upward bids, since no bids are placed in this direction for either hour. In the downward direction, however, the relative variation of the bids is 7.23\% for hour 19 and 12.84\% for hour 10. The larger relative variation of the bids at hour 19 may not correspond directly to the lower relative variation of the empirical percentile of energy flexibility at this hour, but could instead reflect the combined relative variation of the empirical percentiles across all flexibilities, which range between 1.4\% and 4.4\%.


One of our main findings is that the analytical reformulation leads to fewer out-of-sample constraint violations compared to the sample-based approach in almost all cases. For instance, the decrease in constraint violation rate for hour 18 is 6.78\%. This can be attributed to the higher bids made by the sample-based method, which compensates for the greater uncertainty associated with the sampled data. In addition to being more computationally tractable, the analytically formulated problem is also faster to solve, likely due to the significant reduction in the number of constraints. For instance, using 216 samples in the analysis results in approximately 680 constraints for the sample-based method. In contrast, the analytical method uses only 22 of the 216 samples for distribution fitting, and the optimization problem involves just three constraints.

We  calculate the resulting revenue of the EV aggregator from offering ancillary services using historical FCR-D up and down prices in Denmark, defined as:
\begin{equation}
    \sum_h b^{\uparrow *}_h\pi_h^\uparrow + b^{\downarrow *}_h\pi_h^\downarrow,
\end{equation}
where $\pi_h^{(\cdot)}$ represents the prices of the FCR-D capacity reservations, and $b^{(\cdot)*}_h$ denotes the bid placed at hour $h$ in either direction. Using prices from the early auction of the FCR-D up and down markets for the relevant period, corresponding to the data collection timeframe, and the resulting reserve bids, the total reservation payment amounts to €136,292 for the sample-based method and €105,366 for the analytical method over the entire year. Extending the formula to include penalties for bid violations would provide a more complete picture of the aggregator's profit; however, this is outside the scope of the present paper. Nevertheless, considering the results regarding constraint violations, it is likely that bids determined using the sample-based method may incur higher penalties.

\subsection{Sensitivity analysis}
For the sensitivity analysis, we focus exclusively on the proposed analytical approach, investigating how the optimal reserve capacity bids vary with different values of $\varepsilon$. In particular, we examine the effect of smaller $\varepsilon$, which imposes stricter reliability requirements on the bids.

In \eqref{eq:Bonferroni distributed constraints}, the allowance for failure is distributed evenly across all requirements involved in the chance constraints. We can utilize this formulation to create a constraint that allows for a failure allowance lower than the level used in the distribution fitting scheme. Recall that we are using data from the 10\textsuperscript{th} percentile to fit distributions to each flexibility. This allows us to investigate whether we can find solutions to the bidding strategy with a violation allowance lower than 10\%.
To perform the sensitivity analysis on the failure allowance, we can define the constraints similarly to \eqref{eq:bonferroni opt prob CC1}-\eqref{eq:bonferroni opt prob CC3}, replacing the right-hand side of \eqref{eq:Bonferroni distributed constraints} with an arbitrary $\alpha\in(0.0005, 0.1]$, where $\alpha=\varepsilon/3$  corresponds to the Bonferroni corrected problem \eqref{eq:bonferroni opt prob}. Following the same procedure as in Section \ref{sec:analyticalsolution}, we obtain a formulation of the optimization problem for the bidding strategy where we may vary the input reliability level, below the threshold $\varepsilon$:
\begingroup
\allowdisplaybreaks
\begin{subequations}\label{eq:analyticalformulation_eps}
    \begin{align}
    &\max_{b^\downarrow, b^\uparrow \geq 0} b^\uparrow + b^\downarrow \\
        \text{s.t.}\quad& \nonumber\\
         &0.2b^{\downarrow} + b^{\uparrow} \leq r_{0.1}^\uparrow -\left(-\frac{1}{\kappa^\uparrow}\log\left(\frac{\alpha}{\varepsilon} \right) \right)^{1/\gamma^\uparrow} \\
        &b^{\downarrow} \leq r_{0.1}^\downarrow -\left(-\frac{1}{\kappa^\downarrow}\log\left(\frac{\alpha}{\varepsilon} \right)\right)^{1/\gamma^\downarrow} \label{eq:analyticalformulation_eps_2} \\
        &b^{\downarrow} \leq r_{0.1}^{\rm{E_{20}}} -\left(-\frac{1}{\kappa^{\rm{E}}}\log\left(\frac{\alpha}{\varepsilon} \right)\right)^{1/\gamma^{\rm{E}}}. \label{eq:analyticalformulation_eps_3}
    \end{align} 
\end{subequations}
\endgroup

Using this formulation, we can successfully obtain bids in both directions, with $\alpha$ potentially as low as 0.05\% for some hours, depending on the underlying distributions. Following the sampling procedure \eqref{eq:samplesize} with the current sample set, we would not be able to reduce the constraint violation allowance to less than 6\% \citep{CalafioreCampi} following the sample-based method, which represents a reduction of only 4\%. This highlights the advantage of the analytical reformulation in this type of analysis. 

\begin{figure}
    \centering
    \includegraphics[width=0.8\linewidth]{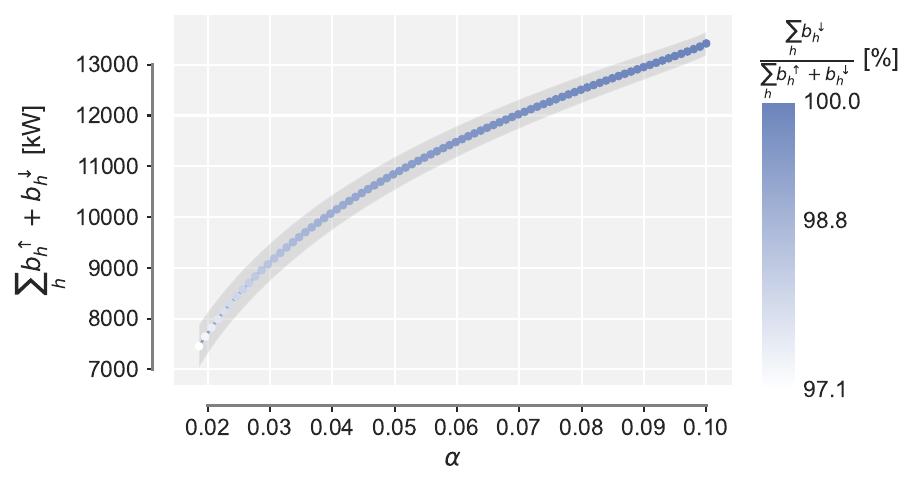}
    \caption{Total bids across 24 hours as a function of $\alpha$ in \eqref{eq:analyticalformulation_eps}, where the dots denote the mean bids over 10 runs, and the gray band signifies the 95\% confidence interval of the sample mean, calculated using \eqref{eq:confint}. The dots are colored based on how much the bid in the downwards direction, $b^\downarrow$, covers the total bid in both directions.}
    \label{fig:total_bid_epsilon}
\end{figure}

Figure \ref{fig:total_bid_epsilon} illustrates the total bids across 24 hours as a function of $\alpha$, as defined in \eqref{eq:analyticalformulation_eps}, where $\alpha=0.02$ is the lowest value for which we could obtain bids for all 24 hours of the day. The bids shown are averages across 10 runs, similar to the results in Figure \ref{fig:results_all}. The point-wise 95\% confidence level is indicated by a grey band, which is calculated in the standard way using the  $t$-distribution \citep{StatBookConfInt}:
\begin{equation}\label{eq:confint}
    CI_{0.95}=\Bigg[\Bar{x}- t_{.025, 9}\frac{s}{\sqrt{10}}, \ \Bar{x} + t_{.025, 9}\frac{s}{\sqrt{10}}\Bigg],
\end{equation}
where $\Bar{x}$ and $s$ represent the sample mean and sample standard deviation, respectively, and $t_{.025, 9}$ is the critical value from the $t$-distribution with 9 degrees of freedom ($df=10-1=9$) for a two-sided 95\% confidence level.

\section{Conclusion and future work}\label{sec:concandfuture}
This paper has developed an analytical bidding model for an electric vehicle aggregator participating in the Nordic ancillary service markets. The proposed model integrates extreme value theory into a chance-constrained optimization framework, effectively addressing the challenge of meeting Energinet’s P90 requirement while reducing computational burden. Compared with a conventional sample-based benchmark, the approach improves out-of-sample reliability by up to 8\% and solves optimization problems up to 4.8 times faster. Furthermore, the methodology successfully generates feasible bids with reliability levels as high as 99.95\%, which would otherwise require prohibitively large scenario sets. These findings demonstrate that incorporating extreme value theory can significantly enhance both robustness and tractability in bidding under uncertainty.

The results also highlight several practical considerations for applying the framework in real-world aggregator operations.

\begin{enumerate}
    \item Data availability: Accurate estimation of flexibility distributions requires high-resolution and reliable data on charging behavior, which may not always be accessible across different fleets.
    \item Operational integration: Aggregators must embed the proposed bidding tool into existing decision-making processes that also consider portfolio management, contractual obligations, and financial risk.
    \item Implementation and scalability: Real-time application calls for efficient software integration with charging platforms and communication systems to ensure daily bid submissions without excessive computational overhead.
    \item Adaptability: Practical deployment may also need to account for temporal correlations in consumption patterns, weekday–weekend variations, and the evolution of charging infrastructure.
\end{enumerate}

This work establishes a regulation-compliant and computationally efficient framework for bidding under uncertainty, offering both methodological innovation and practical insights. Future studies could extend the model by:

\begin{itemize}
    \item incorporating price forecasts, activation data, or dynamic reserve bidding;
    \item testing the approach within real aggregator platforms; and
    \item exploring bidding across multiple markets (e.g., automatic and manual frequency restoration reserves) while considering inter-temporal constraints and dependence structures.
\end{itemize}

Such developments would further enhance the scalability and applicability of the proposed methodology, supporting reliable and cost-efficient participation of electric vehicles in ancillary service markets and strengthening overall power system security.

\begin{appendices}
\section{Flexibility estimation} \label{sec:case study intro}

\begingroup
\allowdisplaybreaks
Our case study utilizes real-world data containing charging profiles from 1400 residential EV charging boxes (CB) in Denmark. The measurements are recorded from March 24, 2022, to March 21, 2023, with an average time-step of 2.84 minutes, which has been interpolated to obtain a one-minute resolution. It is assumed that only one EV is coupled with each CB. The historical EV consumption level serves as the baseline for the estimation of each flexibility.

The acquired data lacks user-specific information, such as the real battery capacity of the EV, the state of charge (SoC), and the maximum power rate of the CB, for which assumptions have been made to calculate them. To estimate flexibility, the data has been manipulated to include the following:
\begin{itemize}
    \item $t$: index for time in minutes, 
    \item $v$: index for EVs, 
    \item $s$: index for charging sessions, 
    \item $k_{v,t}$: a binary parameter indicating whether the EV $v$ is connected to the CB at minute $t$,
    \item $P_{v,t}$: instantaneous power output of the CB to the EV at minute $t$ [kW],
    \item $L_v$: battery capacity of the EV [kWh], 
    \item $P_v^{\max}$: maximum power rate of the CB [kW], 
    \item Current SoC [\%], where it is assumed that the EVs battery is full (100\%) when disconnecting from the CB.
\end{itemize}

We assume that the maximum battery capacity for every EV $v\in V$ is the largest energy output exerted by the CB over \textit{all charging sessions} $s\in S_v$, where $S_v$ denotes the set of sessions for EV $v$. That is, $L_v^{\max}=\max\{\sum_{t} P_{v,t,s}\}$, $t\in[t_{s,0}, t_s^{\max}]$, $t_{s,0}$ and $t_s^{\max}$ being the start and end time of charging session $s$. The SoC at each minute $t$ is given by $L_{v,t}=L_v^{\max} - (L_{v,s}-\sum_{i=1}^t P_{i,v})$, where $L_{v,s}$ is the total energy applied to the vehicle in the current charging session. By this statement, it is assumed that the SoC at the end of all charging sessions is 100\%.

The energy that can be applied to the vehicle without exceeding the battery's capacity is formulated as:
\begin{subequations}
  \begin{align}
   r^{\rm{E}}_{v,t}=(L^{\max}_v-L&_{v,t})\prod_{i=t}^{t+20}k_{v,i},\\ 
\intertext{measured in kWh, where the product over $k_{v,t}$ ensures that the measure is only calculated if the vehicle is continuously connected to the CB in the next 20 minutes. 
To comply with the LER requirement, the constraint is restated:}
   r^{\rm{E}}_{v,t}\geq \frac{20}{60} r^\downarrow \quad&\Rightarrow\quad 3r^{\rm{E}}_{v,t}\geq c^\downarrow, \\
\intertext{where the unit of $3r^{\rm{E}}_{v,t}$ is in kW, signifying how much power the CB could \textit{theoretically} output during the following 20 minutes. There is, however, a \textit{physical} limit on the power output of the CB, and therefore the constraint is further restricted:}
    b^\downarrow \leq \min\{P&^{\max}_{v,t}, 3r^{\rm{E}}_{v,t}\}, 
      \end{align}
\end{subequations}
that is, $r^{\rm{E_{20}}}_{v,t}=\min\{P^{\max}_{v,t}, 3r^{\rm{E}}_{v,t}\}$. 
In the case where the EV is not connected to the charger 20 minutes ahead, it will not have any energy flexibility unless it is considered in a portfolio where another CB can provide the flexibility.

The upwards flexibility, $r^\uparrow$, denotes how much the power applied to the EV can be reduced, and is simply defined as the power applied to the EV $v$ at the current time $t$ given that it is connected to a CB:
\begin{subequations}
  \begin{align}
    r^\uparrow_{v,t} =& P_{v,t}k_{v,t}. \\
\intertext{\indent On the other hand, the downwards flexibility, $r^\downarrow$, denotes how much the power applied to the EV by the CB can be increased, and is the maximum power rate the CB can provide, $P^{\max}_{v}=\max\{P_{v,t}\},t\in T_v$ (all measured time for vehicle), minus the current power rate applied to the connected EV:}
    r^\downarrow_{v,t} = (P^{\max}_{v,t}-&P_{v,t})k_{v,t}. \\
\intertext{\indent Aggregation across vehicles for all flexibilities is simply calculated as the sum over all vehicles' flexibility at each minute $t$:}
    r^{\rm{(\cdot)}}_{t} = \sum_v &r^{\rm{(\cdot)}}_{v,t}, \\
\intertext{where $r^{\rm{(\cdot)}}\in\{r^\uparrow, r^\downarrow, r^{\rm{E_{20}}}\}$. As the reserve bid has to be uniform across each hour $h$, the flexibility is bounded as the minimum available flexibility within the hour:}
    r^{\rm{(\cdot)}}_h = \min_{t\in[t_0, t_{60})}&\left\{r^{\rm{(\cdot)}}_{t}\right\},
  \end{align}
\end{subequations}
\endgroup
$t_0$ and $t_{60}$ being the start and end minute of hour $h$, respectively. Taking the minimum flexibility across the hour in this manner results in a loss of temporal dependencies in the data, which may lead to simplifications in the following analyses. The resulting data contains a year's worth of flexibility for each hour of the day. Illustrative examples of how the flexibilities are calculated are found in Figure \ref{fig:flexcalculations}. In Figure \ref{fig:scatter time} scatter plots of the (minimum) flexibilities for an arbitrary hour over the whole year are shown, where the horizontal line indicates the 10\textsuperscript{th} percentile, which serves as the threshold under which we are selecting data to use for our analysis in the analytical formulation of the bidding strategy.
\newpage
\section{KS test results for distribution fitting validation}

Table \ref{tab:D and p-val} provides the mean and standard deviation of the KS test results over 10 runs.

\begin{table}[h]
    \centering
    \begin{tabular}{|c|cc|cc|cc|}
    \hline
        & \multicolumn{2}{c|}{Downward} & \multicolumn{2}{c|}{Upward} & \multicolumn{2}{c|}{Energy} \\ \hline\hline 
        Hour & $D_n$ (sd) & p-val (sd) & $D_n$ (sd) & p-val (sd) & $D_n$ (sd) & p-val (sd) \\ \hline \hline
        1 & 0.155 (0.023) & 0.943 (0.061) & 0.118 (0.023) & 0.994 (0.005) & 0.127 (0.029) & 0.981 (0.039) \\
        2 & 0.195 (0.043) & 0.776 (0.193) & 0.145 (0.029) & 0.956 (0.058) & 0.141 (0.026) & 0.967 (0.051) \\
        3 & 0.227 (0.048) & 0.621 (0.228) & 0.168 (0.031) & 0.895 (0.109) & 0.182 (0.037) & 0.835 (0.149) \\
        4 & 0.255 (0.038) & 0.496 (0.19) & 0.109 (0.023) & 0.996 (0.005) & 0.227 (0.057) & 0.627 (0.267) \\
        5 & \textcolor{red}{0.273 (0.052)} & \textcolor{red}{0.419 (0.252)} & 0.127 (0.029) & 0.981 (0.039) & 0.25 (0.058) & 0.514 (0.261) \\
        6 & 0.236 (0.047) & 0.591 (0.235) & 0.132 (0.026) & 0.98 (0.038) & 0.205 (0.044) & 0.728 (0.193) \\
        7 & 0.205 (0.039) & 0.74 (0.182) & 0.123 (0.022) & 0.993 (0.005) & 0.191 (0.036) & 0.8 (0.151) \\
        8 & 0.195 (0.031) & 0.788 (0.139) & 0.127 (0.042) & 0.958 (0.115) & 0.173 (0.019) & 0.895 (0.05) \\
        9 & 0.164 (0.023) & 0.919 (0.061) & 0.127 (0.042) & 0.958 (0.115) & 0.155 (0.032) & 0.931 (0.116) \\
        10 & 0.15 (0.037) & 0.932 (0.116) & 0.136 (0.03) & 0.968 (0.051) & 0.132 (0.026) & 0.98 (0.038) \\
        11 & 0.145 (0.019) & 0.966 (0.05) & 0.132 (0.026) & 0.98 (0.038) & 0.145 (0.047) & 0.922 (0.117) \\
        12 & 0.127 (0.019) & 0.992 (0.004) & 0.186 (0.026) & 0.836 (0.113) & 0.127 (0.036) & 0.97 (0.052) \\
        13 & 0.155 (0.032) & 0.932 (0.064) & 0.123 (0.031) & 0.982 (0.039) & 0.132 (0.026) & 0.98 (0.038) \\
        14 & 0.127 (0.029) & 0.981 (0.039) & 0.145 (0.047) & 0.921 (0.152) & 0.132 (0.026) & 0.98 (0.038) \\
        15 & 0.141 (0.026) & 0.967 (0.051) & 0.145 (0.019) & 0.966 (0.05) & 0.136 (0.03) & 0.968 (0.051) \\
        16 & 0.127 (0.029) & 0.981 (0.039) & 0.141 (0.026) & 0.967 (0.051) & 0.145 (0.029) & 0.956 (0.058) \\
        17 & 0.141 (0.026) & 0.967 (0.051) & 0.127 (0.029) & 0.981 (0.039) & 0.127 (0.029) & 0.981 (0.039) \\
        18 & 0.141 (0.014) & 0.978 (0.037) & 0.109 (0.023) & 0.996 (0.005) & 0.109 (0.023) & 0.996 (0.005) \\
        19 & 0.136 (0.03) & 0.968 (0.051) & 0.136 (0.03) & 0.968 (0.051) & \textcolor{blue}{0.105 (0.031)} & \textcolor{blue}{0.986 (0.04)} \\
        20 & 0.127 (0.029) & 0.981 (0.039) & 0.15 (0.022) & 0.955 (0.057) & 0.114 (0.024) & 0.995 (0.005) \\
        21 & 0.114 (0.032) & 0.984 (0.04) & 0.132 (0.026) & 0.98 (0.038) & 0.132 (0.034) & 0.969 (0.052) \\
        22 & 0.132 (0.026) & 0.98 (0.038) & 0.141 (0.014) & 0.978 (0.037) & 0.141 (0.034) & 0.957 (0.059) \\
        23 & 0.173 (0.036) & 0.871 (0.138) & 0.132 (0.026) & 0.98 (0.038) & 0.123 (0.031) & 0.982 (0.039) \\
        24 & 0.145 (0.019) & 0.966 (0.05) & 0.136 (0.0) & 0.99 (0.0) & 0.132 (0.014) & 0.991 (0.003) \\ \hline
    \end{tabular}
    \caption{Mean and standard deviation of the KS test results over 10 runs. The worst values, based on $D_n$, are highlighted in \textcolor{red}{red}, while the best values are highlighted in \textcolor{blue}{blue}.}
    \label{tab:D and p-val}
\end{table}
\end{appendices}


\bibliography{sn-bibliography}


\end{document}